# On the material genome of wurtzite ferroelectrics

Zijian Zhou,[1] Kan-Hao Xue,[1,2,*] Jinhai Huang,[1] Heng Yu,[1] Shengxin Yang,[1] Shujuan Liu,[2] Yiqun Wang,[2] Xiangshui Miao[1,2]

[1]School of Integrated Circuits, Huazhong University of Science and Technology, Wuhan 430074, China.

[2]Hubei Yangtze Laboratory, Wuhan 430205, China

*Corresponding author: xkh@hust.edu.cn

## Abstract

As the dielectric film thickness shrinks to ~10 nm, some traditional wurtzite (WZ) piezoelectric materials demonstrate ferroelectricity through element doping. Among them, Sc doped AlN and Mg doped ZnO are the most famous examples. While it is widely acknowledged that the dopant atoms effectively reduce the coercive field, enabling ferroelectric polarization switching, the material genome of these WZ ferroelectrics is still less understood. In this work, we analyze the features of WZ ferroelectrics, ascribing them to five-coordination (5C) ferroelectrics, which may be compared with 6C ferroelectrics (perovskite-type) and 7C ferroelectrics (hafnia-like). In particular, the exact reason for their adopting the hexagonal WZ structure instead of the zinc blende structure is studied. Emphasis is paid to the degree of ionicity in promoting the hexagonal arrangement, and the phenomenon of layer distance compression is discovered and explained in WZ ferroelectrics. The role of element doping in coercive field reduction is understood within this context.

## Introduction

Ferroelectric random access memory (FeRAM) and ferroelectric field effect transistor have emerged as effective solutions in addressing the current demand on data storage, for a balanced speed and power consumption, as well as its electric voltage-driven nature. The first commercial FeRAM products in the 1990s utilized perovskite ferroelectrics, such as PbZr$_{1-x}$Ti$_x$O$_3$ (PZT), for their sufficient spontaneous polarization and low switching barrier. Nevertheless, perovskite-type FeRAM has long been staying at the 130 nm technical node,[1] for two major reasons. On the one hand, these ferroelectrics require a relatively high crystallization temperature above 600 °C, which is inconsistent with the advanced CMOS technology that usually involves metallization materials such as NiSi$_x$, which can only withstand 450 °C to 500 °C. On the other hand, the vertical scaling performance of typical perovskite ferroelectrics is unsatisfactory, which does not maintain well ferroelectricity at a thickness below 10 nm. The discovery of ferroelectricity in doped hafnia[2] and doped AlN,[3] however, has revived the hope for high-density ferroelectric memory devices. These simple compounds can retain excellent spontaneous polarization characteristics even at scales as small as a few nanometers. Nowadays, hafnia-based ferroelectrics are classified as fluorite-type ferroelectrics,[4,5] while AlN-based ferroelectrics are named wurtzite (WZ) ferroelectrics. Compared with their great application potential, the physical understanding into their origin of ferroelectricity is less mature. Recently, a theory based on chemical coordination numbers has been proposed for the emergence of ferroelectrics in fluorite-type materials,[6] which states that the small cation radius forbids these compounds to place 8 O anions around each cation, reducing the cation coordination number from 8 to 7. An asymmetric way of forming the 7-coordination (7C, and the same short note will be used throughout) configuration leads to the ferroelectric phase. In contrast, WZ ferroelectrics own an even shorter history,[7] and investigation into its ferroelectric material genome has been relatively rare.

Traditional WZ materials are known for their piezoelectric properties, but their ferroelectric performance has been limited by the excessively high coercive field ($E_c$), which typically leads to dielectric breakdown before polarization switching can occur. The problem is alleviated with the advent of Al$_{1-x}$Sc$_x$N with lower $E_c$ values,[3] sparking renewed interest in the ferroelectric properties of WZ structures. Around the same time, doping strategies in other WZ materials have been proposed to induce ferroelectricity, such as Mg-doped ZnO[8,9] and Sc-doped GaN.[10,11] Among AlN-based compounds, replacing Al atoms with elements like Sc[3], Y[12] or B[13,14] has proven effective in lowering the switching barrier. Most explanations attribute these effects to factors like stress or volume changes,[15–17] though they often remain *post hoc* interpretations. Currently, clear theoretical guidance to predict switching barrier behavior faithfully is yet to be established.

The WZ-structured compounds are primarily IIB-VI and IIIA-V compounds. They belong to the $P6_3mc$ space group, where both anions and cations adopt a hexagonal close-packed (*h.c.p.*) configuration. Along the c-axis, there exists a certain displacement between the two hexagonal lattices, which accounts for the possible emergence of ferroelectricity. In *h.c.p.* metals, deviations from the ideal c/a ratio (1.633) are also commonly observed, particularly in Zn and Cd, whose c/a ratios exceed the ideal value and are measured as 1.861 and 1.88, respectively.[18] The study by Zheng-Johansson et al. suggested that the fundamental reason for the abnormal c/a ratio of transition metal elements containing *d*-electrons lies in the difference in the rate of change of the Madelung energy of *d*-electrons when c/a decreases compared to an increased c/a ratio.[18] The abnormal phenomenon of the c/a ratio has also been observed in WZ structures and is regarded as a critical factor in assessing the ferroelectricity of the WZ structure. For example, Hiroshi Yamada et al. identified c/a as a key parameter for the phase transition of the WZ structure and correlated the c/a ratio with the spontaneous polarization intensity of the WZ phase.[19] However, there is no direct evidence to confirm this hypothesis. To explore the relationship between these phenomena, we proceed from the perspective of materials genomics, introducing two reference structures: the zinc blende (ZB) structure with the $F\bar{4}3m$ space group and the hexagonal structure with the $P6_3/mmc$ space group. Both of these structures are centrosymmetric but exhibit high structural similarity to WZ. The ZB structure is another common configuration for IIB-VI and IIIA-V compounds, where anions and cations form a cubic close-packed (*f.c.c*) arrangement with two sets of cubic lattices displaced along the body diagonal. The anions (or cations) are located at the centers of regular tetrahedra formed by the opposite ions, without generating spontaneous polarization. Hence, this study explores the origin of ferroelectricity in WZ structures, focusing on the selection between hexagonal WZ and cubic ZB structures by nature. The generic rule of element doping on switching barrier will also be investigated. And the paper is organized as follows. After introducing the basic settings involved in our calculations, the cation coordination number and lattice constant ratio of the WZ-structure binary compounds are investigated, for a relatively large set of materials. The tendency of their adopting the WZ or the ZB structure will be analyzed. Finally, the effect of doping in coercive field reduction is discussed.

## Computational Details

All first-principles calculations in this study were performed using density functional theory (DFT) under generalized gradient approximation (GGA), as implemented in the Vienna *Ab initio* Simulation Package[20,21] (VASP). The Perdew-Burke-Ernzerhof (PBE) functional[22] was used to account for exchange-correlation interactions. The projector augmented-wave (PAW) method[23,24] was applied, and the valence electrons configurations were: $2s2p$ for B/N/O, $3s3p$ for Al/P/S, $4s3d$ for Zn, $4s4p3d$ for Ga, $4s4p$ for As/Se, $5s4d$ for Cd and $5s5p4d$ for In, $5s5p$ for Sb/Te. A plane-wave cutoff energy of 500 eV was set for all calculations. The $k$-point mesh for relaxation was 11×11×11 centered at Γ. The convergence criteria for ionic relaxation was set at 0.001 eV/Å, and the electronic energy convergence threshold was $10^{-8}$ eV. The climbing image nudged elastic band (CI-NEB) method was employed to compute polarization switching barriers, with force convergence set to 0.01 eV/Å and electronic convergence at $10^{-7}$ eV. Additionally, the shell DFT-1/2 method[25–27] was used to correct self-interaction errors, with inner and outer self-energy potential cutoff radii set at 0.8 and 2.4 bohr, respectively. Lobster software was used to calculate the crystal orbital bond index

(COBI).[28]

**Results and discussion**

Traditional coordination theory identifies the WZ structure as a 4C system, where cations are located within a tetrahedral arrangement of surrounding anions, similar to the ZB phase. Accordingly, it is difficult to understand the difference between WZ and ZB structures from the cation coordination number perspective. The reason lies in that the coordination number has been calculated with the sole reference to the bond lengths. If the WZ structure is considered as 4C, the cation's surroundings can be treated as a tetrahedral environment. During polarization switching, the cation may shuttle between neighboring tetrahedra, which fails to capture the key geometric feature (**Figure 1(a)**). Recently, we have proposed a coordination number theory (mixed length-angle coordination number theory, MLAC) that refers to both the bond lengths and bond angles,[6] and it is shown to be particularly suitable for ferroelectrics. The basic idea is (see Supporting Information **Note 1** for details), based on a central cation, to assign each neighboring anion a finite solid angle. Once the entire space is parceled out, additional anions located farther apart will no longer be counted as neighbors. Quantitatively, a new anion could be included in the neighbors only if its angle to any existing neighboring bond is greater than 65° (the reason for this threshold angle was specified in the original publication[6]). The MLAC theory fits the analysis of bulk (not two-dimensional) ferroelectrics since it takes into consideration the particular anion apart that has the potential to approach the central cation through ferroelectric switching. This coordination theory may be applied to analyze WZ and ZB compounds. In the WZ structure, the $\theta$ angle corresponding to the fifth anion determines whether it adopts a 4C or 5C configuration (**Figure 1(b)**). To this end, we first arranged all IIB-VI and IIIA-V compounds to be in the (possibly hypothetical) WZ structure. After performing full structural relaxation within the hexagonal framework, the geometric data are given in **Figure 1(c)**. Among them (see data listed in **Table S1**), the one that is closest to 4C is BP. Manually set in the WZ structure, BP has the smallest fifth anion bond angle, $\theta =$ 70.14°, but it is still larger than the critical 65° angle, meaning that the fifth P anion is still classified as a coordinated anion. This does not mean BP should adopt the 5C WZ structure in reality; it is naturally in the ZB structure unless constrained in the WZ configuration as in our research. To sum up, the WZ structure could be considered a 5C system[29] according to the MLAC theory. In this configuration, the anion center establishes the base of a trigonal bipyramid, while the cation is located at the center of the upper tetrahedron. During the polarization switching, the "shuttle" phenomenon observed in the 4C configuration is avoided.

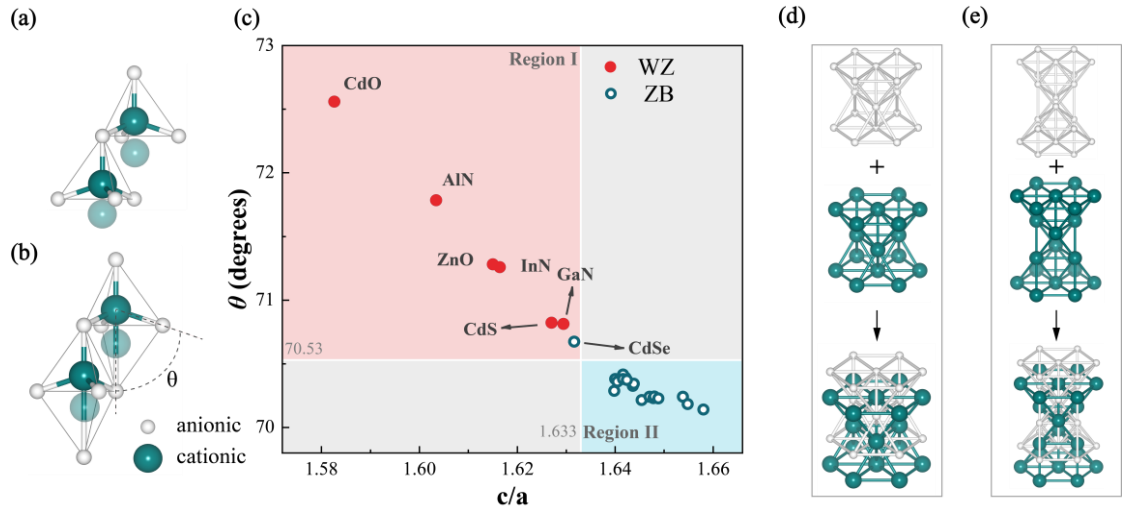

**Figure 1.** The polyhedron formed by nearest neighbor atoms in traditional coordination theory (a) and MLAC theory (b). The translucent cation represents the position of the cation after polarization switch. (c) The relationship between the structural state and its distortion. The entire diagram is divided into four regions by two reference lines: the $N_5$ (*cf.* **Figure N1**) bond angle of a regular tetrahedron (70.53°) and the ideal $c/a$ ratio ($\sqrt{8/3} = 1.633$) for a regular tetrahedron. Region-I indicates *c*-axis compression, while Region-II indicates *c*-axis elongation. The closer a point is to the top-left corner, the stronger the compression effect in the system. (d) WZ structure. (e) ZB structure.

Since all WZ structures in our research are 5C, this fact can be acknowledged for the discussions below. According to the statistical data in **Figure 1(c)**, the critical angle $\theta$ and the c/a ratio show a clear linear relationship, indicating some consistent physical significance. Thus, we focus on $\theta$ due to its special relationship with the coordination number, and as a continuously-varying parameter, it owns a finer resolution compared with the coordination number. In **Figure 1(c)**, almost all materials are distributed across Region-I and Region-II, with materials that have a ZB ground state predominantly located in the Region-II (except for CdSe). These materials, when arranged in the WZ structure, tend to show a stretched state (*i.e.*, $c/a > 1.633$), which is less stable than the ZB structure from an energy perspective. In contrast, materials with a WZ ground state are exclusively found in the Region-I, indicating that only under *c*-axis compression can the WZ structure maintain a lower energy than the ZB structure. Naturally, this leads to the question as to why WZ materials require a compressed structure to remain energetically favorable.

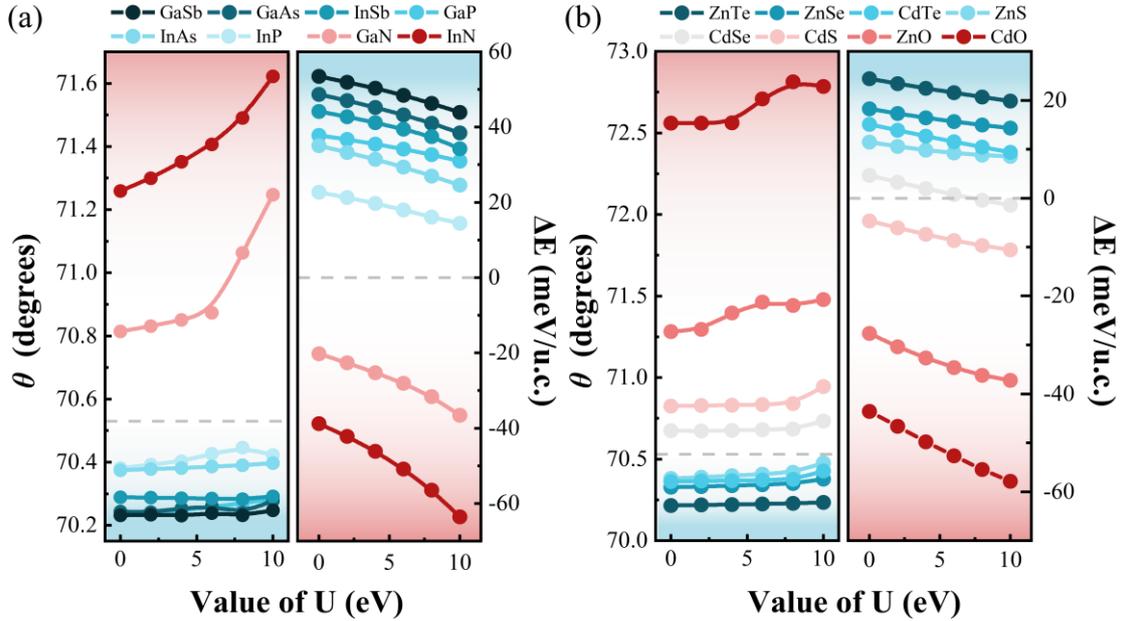

**Figure 2.** Energy change $\Delta E = E_{\mathrm{WZ}} - E_{\mathrm{ZB}}$ and $\theta$ corresponding to different $U$ values after DFT+U treatment： (a) III-V, (b) II-V. DFT+U method was performed on the *d*-electrons of cations only, with $J = 0$. A smaller value $\Delta E$ hints at a more stable WZ structure against the ZB structure.

We speculate that the ionic character of the system is the predominant factor contributing to the compression of WZ. There are two reasons for this. On the one hand, all WZ materials under investigation are oxides and nitrides, characterized by a large electronegativity difference between their cations and anions. As shown in **Figure 1 (c)**, materials located in the upper left corner exhibit higher ionic character compared to those in the lower right corner. The two materials in the uppermost left corner are CdO and AlN, while those in the lower right corner, such as BP, BAs, and BN, are strongly covalent compounds. On the other hand, based on previous studies, the ferroelectricity of doped hafnia and perovskite ferroelectric materials is related to their indispensable covalent component in bonding.[6,30,31] Similarly, in WZ structures, ferroelectricity may also originate from covalent bonds, while an increase in ionicity could weaken the ferroelectricity.

We expect to observe changes in the crystal structure by tuning the ionicity of the material. Although it is challenging to experimentally provide a suitable method to explore the impact of ionic or covalent character, first-principles calculations permit us to investigate this issue by altering atomic properties in a hypothetical manner. The Hubbard U correction (DFT+U) can theoretically modify the ionic character of a system to a certain degree by lowering the energy levels of occupied states and raising those of unoccupied states. For IIB-V and IIIA-V materials, the metallic cations generally possess fully occupied $d$-orbitals, except for Al and B, which lacks $d$-electrons. By applying this method to the $d$-electrons of the metallic cations, the contribution of these electrons to the valence band can be reduced, thereby decreasing their role in bonding. The integrated crystal orbital bond index (ICOBI), as proposed by Müller et al.,[28] serves as a tool to evaluate ionicity and covalency: values close to 0 indicate strong ionicity, while values close to 1 reflect strong covalency. Several representative materials (WZ-ZnO, WZ-ZnS, WZ-GaN, WZ-GaAs) were selected as references, and the ICOBI under various $U$ values were calculated **(Table S2)**. The results indicate that ICOBI tends to decrease after processing with DFT+U, confirming an enhancement in ionicity. Furthermore, the energy difference $\Delta E$ between the WZ and ZB structures and the $\theta$ values under these $U$ values were also calculated for all materials containing metallic $d$-electrons **(Figure 2)**. It is noteworthy that $\Delta E$ and $\theta$ would decrease and increase upon enhancing the value of $U$, respectively. Based on these findings, three conclusions can be drawn:

i. The ionic character is indeed positively correlated with $\theta$.

ii. The presence of $d$-electrons is unfavorable for the emergence of the WZ structure.

iii. The presence of $d$-electrons contributes to a decreasing $\theta$.

The last point can potentially be explained by the theory proposed by Jang et al.. As the energy level of the $d$-electrons decreases, the hybridization between the $d_{z^2}$ orbital of the cation and the $p_z$ orbital of the anion gets weakened.[32] This reduces the asymmetry of the wave function along the $c$-axis, causing the cation to shift closer to the center of the trigonal bipyramid, thereby increasing the $\theta$ value. However, this theory is not applicable to AlN as it lacks $d$-electrons.

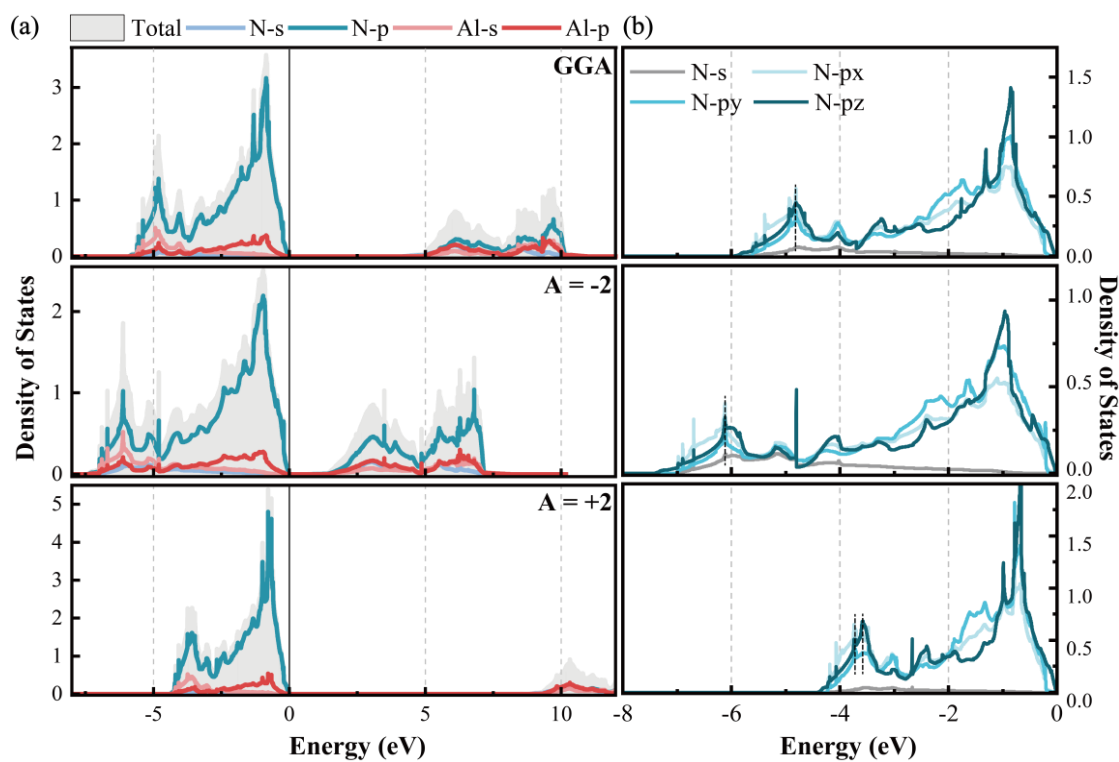

**Figure 3.** (a) PDOS of AlN, calculated with shell DFT+A-1/2 at various A values. (b) DOS for the $p$-orbital electrons components and the $s$-orbital electrons of N, calculated with shell DFT+A-1/2 at various A values.

To validate the previous hypothesis that the $\theta$ angle is highly relevant to the ionicity,

it is necessary to study AlN independently. The first step is to understand how the $\theta$ angle of WZ-AlN behaves with respect to the changes in ionicity. **Figure 3(a)** presents the projected density of states (PDOS) of WZ-AlN calculated through conventional GGA. Near the Fermi level ($E_f$), the Al-$p$ orbitals overlap with the N-$p$ orbitals, exhibiting similar peak locations, which indicates covalency. However, the contribution of N-$p$ electrons is several times higher than that of Al-$p$ electrons, suggesting that the strong electronegativity of N and the metallic nature of Al lead to a significant ionic character in the system. In particular, the DFT-1/2 method introduces the so-called self-energy potentials in solids,[26] which rectifies the spurious electron self-interaction that usually exists in local density approximation (LDA) or GGA calculations. For covalent compounds, the method is extended to shell DFT-1/2.[25] While these methods aim at predicting rectified band gaps, they may also be utilized to perturb the electronegativity of specified atoms,[33] where the strong self-energy potential is usually modulated by a factor $A$ (denoted as "+$A$"). Here we employed the shell DFT+A-1/2 algorithm to study the influence of chemical bonding on the geometric structure. It modifies the pseudopotential of the specified atomic shell, tuning the strength of the self-energy potential calculation in that region, which adjusts the electronegativity of the specified atom and further affects its bonding state. When the $A$ value is less than zero, a reduction in electronegativity occurs; at $A = 0$, the original GGA calculation is retained; and with a positive $A$ value, the electronegativity of the specified atom is increased. The greater the absolute value of $A$, the more significant the modification of electronegativity.[33] Hence, the objective is to manually set the Al-N bonds more ionic, or more covalent, in order to explore the consequences.

Following this approach, we have manually adjusted the electronegativity of the N atom. The PDOS was calculated for various $A$ values using the shell DFT+A-1/2 method based on the GGA-relaxed WZ-AlN structure (**Figure 3(a)**). When $A$ is set to $-2$, the N-$p$ electrons transfer from the valence band to the conduction band, resulting in a decreased electron density within the valence band, while the electronic state of Al remains essentially unchanged. Consequently, the proportion of Al electrons contribution within the valence band increases appreciably, suggesting an enhancement in the covalent character of the system, which validates the effectiveness of this method. When $A$ is set to $+2$, its effect is opposite to that of $A = -2$, with a decrease in the Al electrons proportion within the valence band, indicating an enhancement of the ionic character. Notably, at the second valence band peak near $E_f$, the change in the proportion of Al electrons is more pronounced. For the system after self-energy potential perturbation, structural relaxation was performed under the $P6_3mc$ space group. The $\theta$ value exhibits a clear linear relationship with $A$ (**Figure 5(a)**), demonstrating that the effect of ionic character on WZ-AlN is consistent with other systems containing $d$-electrons.

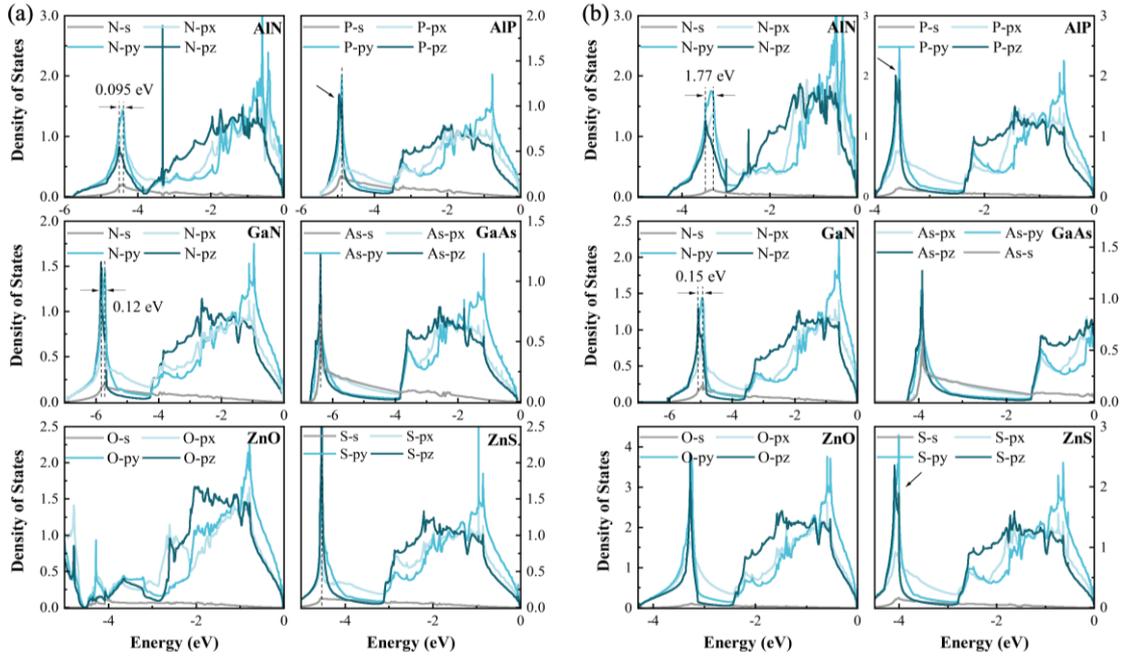

**Figure 4.** DOS for the $p$-orbital electrons components and the $s$-orbital electrons of ZB-AlN, ZB-AlP, ZB-GaN, ZB-GaAs, ZB-ZnO, and ZB-ZnS before (a) and after (b) treatment with the shell DFT+A-1/2 ($A = +2$) or DFT+U ($U = 10$ eV, $J = 0$) method.

Regardless of whether containing $d$-electrons, the variation of c/a with ionic character remains consistent. Therefore, the influence on c/a from an electronic structure perspective is more likely attributed to $s$- and $p$-electrons. Focusing on the electronic structures of ZB and WZ, **Figure 4(a)** illustrates the density of states (DOS) for the *p*-orbital electrons components and the *s*-orbital electrons of anions in ZB-AlN, ZB-AlP, ZB-GaN, ZB-GaAs, ZB-ZnO, and ZB-ZnS. As the most representative ZB material, GaAs exhibits the most typical electronic DOS for anions. Around $-6.4$ eV, the *s*-orbitals of As and the three sub-orbitals of *p*-electrons nearly completely overlap, demonstrating strong $sp^3$ hybridization. This hybridization effect stabilizes the anion at the tetrahedral center of the cation, ensuring sufficient stability of the ZB structure. Similar to ZB-GaAs, ZB-AlP and ZB-ZnS also exhibit strong $sp^3$ hybridization at deeper energy levels. Although the proportion of *s*-orbitals in their anions is relatively low, it is evident that the peaks of their *s*-orbitals align with those of the three *p*-electron sub-orbitals at the same energy level. For ZB-AlN and ZB-GaN, in their WZ ground state, the *s*- and *p*-electrons exhibit similar shapes at deeper energy levels, indicating a certain degree of $sp^3$ hybridization. The difference lies in that, the $p_z$ peak is offset from the peaks of the other orbitals. In other words, ZB-AlN and ZB-GaN show both a tendency for $sp^2$ hybridization and a degree of $sp^3$ hybridization. This dual behavior preserves the tetrahedral structure while causing the cation to shift from the center of the tetrahedron, as $sp^2$ hybridization primarily occurs within the plane. ZB-ZnO is an exception, as its DOS shape differs from other materials. This discrepancy arises from anomalies in the electronic structure of ZB-ZnO calculated by PBE, where the Zn-*d* electrons dominate around -4 eV, disrupting the electronic density distribution of O (**Figure S1**).

The DOS obtained from shell DFT+A-1/2 ($A > 0$) or DFT+U treatments reinforces this hypothesis. After these treatments, the structural privilege of the ZB structure shows a declining trend, and the $sp^3$ hybridization in materials with a ZB ground state is weakened. The DOS reveals different behaviors across materials (**Figure 4(b)**). At the overlap region of *s*- and *p*-electrons, the proportion of *s*-electrons decreases in ZB-GaAs, the $p_z$ peak of ZB-ZnS splits with a reduced overlap proportion, and the original $p_z$ peak of ZB-AlP exhibits even more pronounced splitting. For materials with a WZ ground state, the tendency for $sp^2$ hybridization becomes more remarkable. The offset of the $p_z$ peak relative to other orbital peaks increases in ZB-AlN and ZB-GaN, changing from the original 0.095 eV and 0.120 eV to 0.177 eV and 0.150 eV, respectively, after the treatment. Considering that both shell DFT+A-1/2 ($A > 0$) and DFT+U narrow the valence band, the increase in the offset is even more significant in practice. A similar $p_z$ offset is also observed in ZB-ZnO after the treatment. This offset phenomenon is also observed in WZ-AlN (**Figure 3(b)**). When $A$ is set to $+2$, the offset of the $p_z$ orbital relative to other orbitals increases, while for $A = -2$, the $p_z$ offset disappears. **Figure 5(b)** presents the DOS of $P6_3/mmc$-AlN, where the $p_x$ and $p_y$ orbitals exhibit similar orbital shapes, while the shape of the $p_z$ orbital differs from them. In the deeper valence band, the $p_z$ orbital displays a displacement relative to $p_x$ and $p_y$, reflecting strong $sp^2$ hybridization between N-$s$ and N-$p$ orbitals, which corresponds to its layered structure. The offset observed in $P6_3/mmc$-AlN closely resembles the DOS of WZ-AlN processed with shell DFT+A-1/2 ($A > 0$), confirming that an increase in ionicity indeed drives the structural transition of WZ-AlN towards a $P6_3/mmc$-AlN configuration, promoting an increase in $\theta$.

Within the understanding of the electronic structure, it is now feasible to discuss the origin of the distinct geometric structures in ZB, WZ, and $P6_3/mmc$ phases. The ZB structure, when viewed along the (111) direction, is observed as a *f.c.c* lattice where cations and anions alternate in an ABCABC… stacking sequence (**Figure 1(e)**). In this arrangement, the displacement along the (111) direction places each anion (or cation) precisely at the tetrahedral center of the other type of ions. The WZ structure can be considered a transitional structure between the ZB and $P6_3/mmc$, with ionic character acting as a parameter that adjusts the tendency of WZ towards either extreme. The fundamental driving factor in this transition is the competition between $sp^2$ and $sp^3$ hybridization. When the ionic character is weak, $sp^3$ hybridization dominates, causing the cation lattice to shift upward (assuming the anion lattice is fixed), during which $\theta$ decreases (**Figure 5(c)**). When the cation reaches the center of the anion tetrahedron, the ZB structure, with its higher symmetry, becomes more stable than the WZ structure, and the material adopts the ZB configuration. Conversely, when the ionic character is strong, $sp^2$ hybridization becomes dominant, driving the cation lattice downward until the $P6_3/mmc$ structure is formed, during which $\theta$ increases. From the perspective of MLAC theory, the relationship between ionic character and $\theta$ is intuitive. A larger $\theta$ value corresponds to more spatial freedom for the final coordinating atom, allowing for a higher coordination number. A higher coordination number, in turn, signifies stronger ionic character.

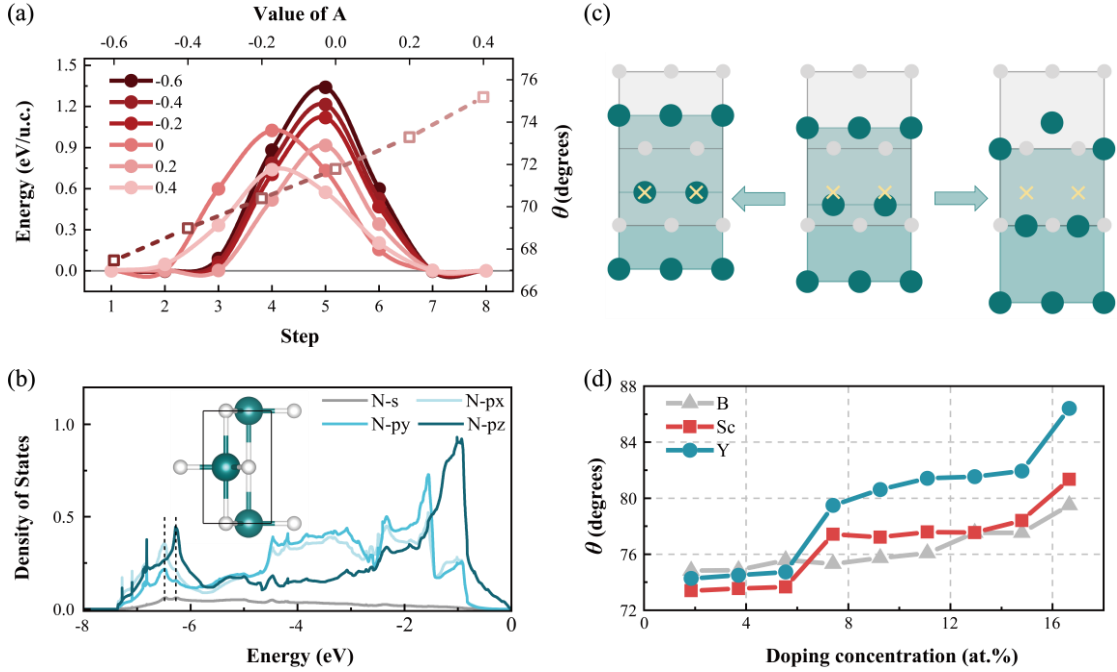

**Figure 5.** (a) CI-NEB results and the angle $\theta$ for WZ-AlN with different $A$ values. (b) Schematic diagram of the relationship between hexagonal lattice movement and structure, where green represents cations, white represents anions, and × denotes the anion tetrahedral center. (c) DOS for the p-orbital electrons components and the s-orbital electrons of $P6_3/mmc$-AlN. (d) Relationship between $\theta$ and doping concentration in WZ-AlN with a 3×3×3 supercell (54 Al atoms in total).

It follows that the enhancement of ionicity can reduce $E_c$ in WZ-compounds, thus enabling the possible ferroelectricity. The $P6_3/mmc$ structure closely resembles the intermediate structure during the polarization switching of the WZ phase. The enhancement of ionicity leads to an increase in $\theta$, causing the polarized WZ structure to more closely resemble the $P6_3/mmc$ structure, which shortens the transition process from the polarized state to the intermediate state, thereby reducing the switching energy barrier. This observation is supported by computational results (**Figure 5(a)**). This theory can explain the mechanism by which doping reduces the energy barrier. **Figure 5(d)** presents the $\theta$ values for AlN doped with Sc, Y and B. As the doping concentration increases, $\theta$ increases, and the $E_c$ decreases for Sc and Y. From another perspective, ScN takes the 6C rock salt structure while AlN is in 5C WZ structure, confirming the stronger ionic tendency of Sc compared with Al. Therefore, the decrease in $E_c$ caused by doping Sc and Y results from the enhanced ionic character of the system. However, B doping does not respect this trend, likely due to its different switching path, where the intermediate state no longer belongs to the $P6_3mc$ space group.

## Conclusion

While the switching dynamics of WZ ferroelectrics are believed to be much simpler than hafnia-based ferroelectrics, there is still a critical question for the origin of the hexagonal WZ structure in compounds like AlN, against the ZB structure. In contrary to the traditional way of coordination number identification, which lists ZB and WZ compounds as 4C compounds, the newly proposed MLAC definition of coordination number identifies the WZ structure as in a 5C state for the cation. The increased coordination number is related to the higher ionicity in WZ compounds compared with ZB compounds. The competition between $sp^2$ and $sp^3$ hybridization is the fundamental cause of the compression in the WZ structure. A larger $\theta$ provides more space for the fifth anion, stabilizing the 5C structure. Additionally, an increased $\theta$ brings the polarized structure closer to the intermediate state of polarization switching, thereby lowering the switching barrier. To reduce $E_c$ through doping, one can replace cations with elements of higher metallicity or anions with elements of higher electronegativity.

# Supplementary Information for

# On the material genome of wurtzite ferroelectrics

Zijian Zhou,[1] Kan-Hao Xue,[1,2,*] Jinhai Huang,[1] Heng Yu,[1] Shengxin Yang,[1] Shujuan Liu,[2] Yiqun Wang,[2] Xiangshui Miao[1,2]

[1]School of Integrated Circuits, Huazhong University of Science and Technology, Wuhan 430074, China.

[2]Hubei Yangtze Laboratory, Wuhan 430205, China

*Corresponding author: xkh@hust.edu.cn

## Note 1. Mixed length-angle coordination theory

The coordination number theory based a mixed consideration of bond length and bond angles is explained in **Figure N1**. Based on the cation in a binary compound, for example, it counts the number of anions surrounding the cation that can be considered as forming bonds with it.

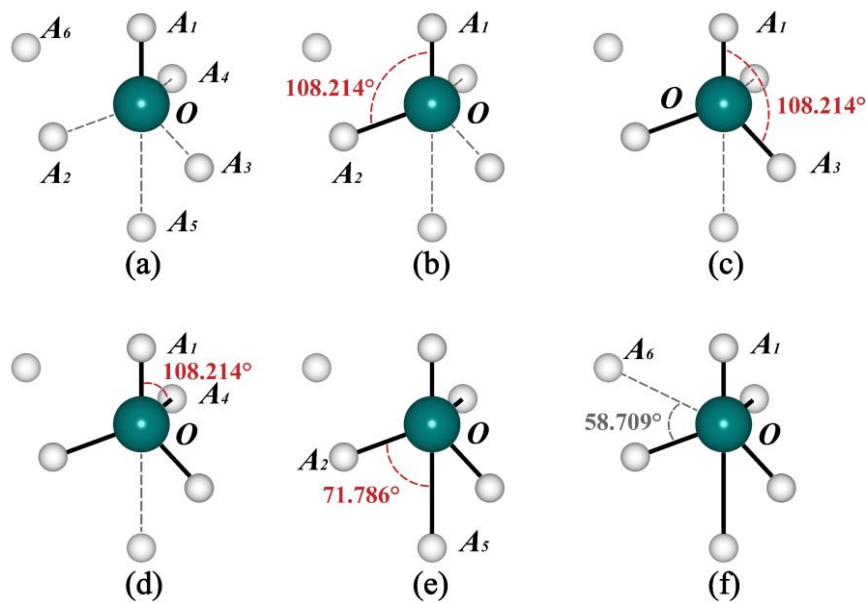

**Figure N1.** Schematic diagram of coordination number calculation.

a) The target atom is labeled as O, and the surrounding atoms are labeled $A_1$, $A_2$, $A_3$, etc., in order of increasing distance from the target atom.

b) The bond angle $A_2$-O-$A_1$ is calculated. If this angle is greater than 65°, $A_2$ is included in the coordination number.

c) The bond angles between $A_3$ and the atoms already included in the coordination number, such as $A_3$-O-$A_1$ and $A_3$-O-$A_2$, are calculated. The smallest angle is selected, and if it is greater than 65°, $A_3$ is included in the coordination number.

d) This process is repeated iteratively until the smallest bond angle involving $A_n$ is less than 65°, at which point the coordination number is determined to be n-1.

**Figure N1** illustrates a 5C system calculated using this theory, where the minimum bond angles for $A_1$ through $A_5$ are all greater than 65°, and thus are included in the coordination number. However, $A_6$ is excluded, as its minimum bond angle is 58.709°, which is less than 65°.

**Table S1.** The $c/a$ axial ratio and $\theta$ angle of all IIIA-V and IIB-VI compounds arranged in the WZ structure.

| IIIA-V | $c/a$ | $\theta$ (degree) | IIB-VI | $c/a$ | $\theta$ (degree) |
|---|---|---|---|---|---|
| BN | 1.6539 | 70.2426 | ZnO | 1.6150 | 71.2822 |
| BP | 1.6582 | 70.1412 | ZnS | 1.6400 | 70.3810 |
| BAs | 1.6548 | 70.1846 | ZnSe | 1.6436 | 70.2558 |
| AlN | 1.6034 | 71.7858 | ZnTe | 1.6453 | 70.2135 |
| AlP | 1.6416 | 70.4171 | CdO | 1.5826 | 72.5592 |
| AlAs | 1.6437 | 70.3446 | CdS | 1.6269 | 70.8242 |
| AlSb | 1.6469 | 70.2412 | CdSe | 1.6315 | 70.6757 |
| GaN | 1.6293 | 70.8139 | CdTe | 1.6404 | 70.3614 |
| GaP | 1.6478 | 70.2316 | | | |
| GaAs | 1.6479 | 70.2438 | | | |
| GaSb | 1.6488 | 70.2291 | | | |
| InN | 1.6163 | 71.2593 | | | |
| InP | 1.6417 | 70.3811 | | | |
| InAs | 1.6423 | 70.3815 | | | |
| InSb | 1.6400 | 70.4237 | | | |

**Table S2**. The integral Crystal Orbital Bond Index (ICOBI) of WZ-ZnO, WZ-ZnS, WZ-GaN, and WZ-GaAs.

|     | GGA     | DFT+U ($U = 10$ eV) |
|-----|---------|---------------------|
| **ZnO**  | 0.18587 | 0.18415 |
| **ZnS**  | 0.10664 | 0.10372 |
| **GaN**  | 0.65582 | 0.62527 |
| **GaAs** | 0.82791 | 0.81132 |

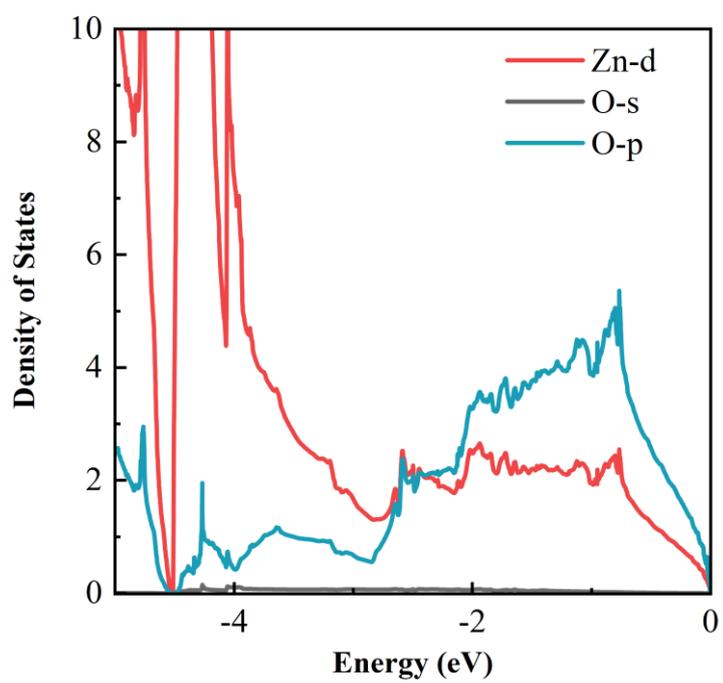

**Figure S1.** PDOS of WZ-ZnO near $E_f$.